\documentclass[twocolumn,10pt,prl,aps,showpacs]{revtex4}

\def\3nab{\tilde{\nabla}}

\def\be {\begin{equation}}
\def\ee {\end{equation}}
\def\bea {\begin{eqnarray}}
\def\eea {\end{eqnarray}}

\newcommand{\sfrac}[2]{{\textstyle{#1\over#2}}}
\def\case#1/#2{\textstyle\frac{#1}{#2}}
\usepackage{graphicx}
\usepackage{dcolumn}
\usepackage{bm}
\usepackage{longtable}

\def\cqg{{\em Class. Quantum Grav.\/} }
\def\grg{{\em Gen. Rel. Grav.\/} }
\def\prd{{\em Phys. Rev.\/} {\bf D}}

\def\aph{{\em Ann. Phys. (NY)\/} }
\def\plb{{\em Phys. Lett.\/} {\bf B}}

\begin{document}

\title{On the co--existence of matter dominated and accelerating solutions in $f(\mathcal{G})$--gravity}

\author{ Naureen Goheer${}^1$,
Rituparno Goswami${}^{1}$, Peter K. S. Dunsby${}^{1,2}$ and Kishore Ananda${}^1$}

\affiliation{1. Department of Mathematics and Applied Mathematics,
  University of Cape Town, 7701 Rondebosch, Cape Town, South Africa}

\affiliation{2. South African Astronomical Observatory,
  Observatory 7925, Cape Town, South Africa.}

\date{\today}

\begin{abstract}

Working within the theory of modified Gauss-Bonnet gravity, we show that FLRW--like  power--law solutions only exist for 
a very special class of  $f(\mathcal{G})$ theories. Furthermore, we point out that any transition from decelerated to accelerated
expansion must pass through $\mathcal{G}=0$, and no function $f(\mathcal{G})$ that is differentiable at this point can admit both a
decelerating power--law solution and any accelerating solution. This strongly constrains the cosmological viability of $f(\mathcal{G})$--gravity,
since it may not be possible to obtain an expansion history of the universe which is compatible with observations.  We explain why the same issue does not occur in $f(R)$--gravity and discuss possible caveats for the case of $f(\mathcal{G})$--gravity.

\end{abstract}

\pacs{98.80.Cq}
\maketitle

\section{Introduction}

One of the most remarkable developments in cosmology over the past decade has been the confrontation of the standard
model of cosmology with observations, leading to the conclusion that the Universe is accelerating in the current epoch.
This rather counter-intuitive result has led to one of the most challenging theoretical puzzles in 21st century physics, namely
to provide an explanation for this late-time cosmic acceleration.  Currently the most popular idea is that the energy density
of the Universe is at the present time dominated by some mysterious dark component known as dark energy and there have been many
proposals offered to explain its nature. At the moment, the one which appears to fit all available observations (Supernovae
Ia \cite{sneIa}, Cosmic Microwave Background Anisotropies \cite{cmbr}, Large Scale Structure formation \cite{lss},
baryon oscillations \cite{baryon}, weak lensing \cite{wl}), turns out to be the so called {\it Concordance Model}
in which a tiny cosmological constant is present \cite{astier} and ordinary matter is dominated by Cold Dark Matter (CDM). However,
despite its success, the $\Lambda$\,-\,CDM model is affected by significant fine-tuning problems related to the vacuum energy scale, so
other exotic negative-pressure fluids, often described in terms of scalar fields \cite{dark}, have been proposed to address these
issues. The problem remains however, that at the present time, there is no direct experimental evidence for the existence of the
scalar fields responsible for the late time (and also the early time) accelerated expansion rate of the Universe and this
has led to a search for other viable theoretical schemes, many of which are based on the idea that the "dark sector" originates
from modifications of the gravitational interaction itself.

Currently, one of the most popular alternatives to the {\it Concordance Model} is based on modifications of the
Einstein-Hilbert action.  Such models first became popular in the 1980's because it was shown that they naturally
admit a phase of accelerated expansion which could be associated with an early universe inflationary phase \cite{star80}.
The fact that the phenomenology of Dark Energy requires the presence of a similar phase (although
only a late time one) has recently revived interest in these models. In particular, the idea that Dark Energy may have a
geometrical origin, i.e., that there is a connection between Dark Energy  and a non-standard behavior of gravitation on
cosmological scales is currently a very active area of research. Additionally, such modifications are appealing due to the 
fact that they can evade constraints on the strength of the gravitational field which are restricted to below solar system scales.

One of the most promising modifications to date are those based on gravitational actions which are
non-linear in the Ricci curvature $R$ and$/$or contain terms involving combinations of derivatives of
$R$ \cite{DEfR,kerner,teyssandier,magnanoff}. These theories  have provided a number of very interesting
results on both cosmological  \cite{ccct-ijmpd,review,cct-jcap,otha,perts} and astrophysical
\cite{cct-jcap,cct-mnras} scales. One of the nice features of these theories is that
the field equations can be recast in a way that the higher order corrections are written as an
energy\,-\,momentum tensor of geometrical origin describing an ``effective" source term on the
right hand side of the standard Einstein field equations  \cite{ccct-ijmpd,review}. In this
{\em Curvature Quintessence} scenario, the cosmic acceleration can be shown to result from such a
new geometrical contribution to the cosmic energy density budget,  due to higher order corrections of the
Hilbert-Einstein Lagrangian.

More recently modifications based on a Lagrangian density which is some general function of the Ricci scalar and the
Gauss-Bonnet term $f(R,\mathcal{G})$ \cite{Sean,Nojiri} have been studied in the context of cosmology. In particular a simple
extension of Einstein gravity: $f(R,\mathcal{G})=R/2+f(\mathcal{G})$ has been investigated by a number of authors \cite{Nojiri-Odin,deFelice,Li-Barrow,Davis,Nojiri:2007bt}. Corrections of this type can be motivated from low-energy string effective actions and compactification of other higher dimensional theories. 
Uddin {\it et al.} \cite{Lidsey} considered  the existence and stability of power-law scaling solutions in $f(\mathcal{G})$ models and found that scaling 
solutions exist in the model $f(\mathcal{G})=\pm 2\sqrt{\alpha \mathcal{G}}$, where $\alpha$  is an arbitrary constant. Inspired by the work on $f(R)$ gravity by Carloni {\it et. al.} \cite{Carloni} and Amendola {\it et. al.} \cite {Amendola}, Zhou {\it et. al.}  \cite{Zhou} used dynamical systems techniques to study the cosmology of $f(\mathcal{G})$ models and argued that one could find cosmologically viable trajectories that mimic the $\Lambda$CDM cosmic history in the radiation and matter dominated  periods, but also have a distinctive signature
at late times

In this paper, we investigate the cosmological viability of $f(\mathcal{G})$ models by investigating the conditions under which one can find
power-law solutions that mimic the standard FLRW expansion history of the Universe. We discover that such solutions only
exist for a very special class of $f(\mathcal{G})$ theories. Furthermore, we show that for there to be a transition between decelerated
and accelerated expansion phases, $\mathcal{G}$ must pass through zero, and there are no functions $f(\mathcal{G})$
that are differentiable at this point that admit both a power-law decelerating solution and any accelerating solution. 
This seriously constrains the cosmological viability of $f(\mathcal{G})$--gravity, since it may not be possible to find an expansion
history of the Universe compatible with observations.

\section{Field equations for general $f(\mathcal{G})$}
We consider the following action within the context of four dimensional
homogeneous and isotropic spacetimes i.e., the Friedmann--Lema\^{\i}tre--Robertson--
Walker (FLRW) universes with no spatial curvature,
\begin{equation}\label{lagr f(R)}
\mathcal{A}=\int d^4 x \sqrt{-g}\left[R +f(\mathcal{G})+{\cal L}_{m}\right]\;,
\end{equation}
where $R$ is the Ricci scalar, and $f$ is general differentiable (at least $C^2$)
function of the Gauss-Bonnet term,
\begin{equation}
\mathcal{G}=R^2-4R_{\alpha\beta}R^{\alpha\beta}+R_{\alpha\beta\mu\nu}R^{\alpha\beta\mu\nu}\;,
\end{equation}
and $\mathcal{L}_m$ represents the matter contribution. As we know, in four dimensions,
the Gauss-Bonnet term is a total differential, and hence for  $f(\mathcal{G})=\mathcal{G}$,
the field equations remain invariant. However for other functions this term has non-trivial
contributions to the field equations. For the homogeneous and isotropic spacetimes, the field equations are:
\begin{itemize}
\item
The Raychaudhuri equation
\begin{eqnarray}
\dot{\Theta}+\frac{1}{3}\Theta^2 =-\frac{\kappa^2}{2}\left(\rho +3P\right)
+\frac{4}{9}\Theta^3 f_{\cal{G}\cal{G}}\dot{\mathcal{G}} \nonumber\\
-\left(f -\mathcal{G} f_{\mathcal{G}}\right)
-\frac{3}{\Theta} \mathcal{G} \dot{f}_{ \mathcal{G} } -\frac{4}{3}\Theta^2
\ddot{f_{\mathcal{G}}},
\end{eqnarray}
where $\Theta$ is the volume expansion which defines a scale factor
$a(t)$ along the fluid flow lines via the standard relation
$\Theta=3\dot{a}/{a}$, and  $f_{n\mathcal{G}}$ abbreviates $\partial^n f/{(\partial \mathcal{G})^n}$ for $n=1,2$,
\item
The Friedmann equation
\begin{equation}\label{fried}
\Theta^2+\frac{8}{3}\Theta^3 f_{\cal{G}\cal{G}}\dot{\mathcal{G}}+3f=
3\kappa^2 \rho+3\mathcal{G}{f}_{\mathcal{G}},
\end{equation}
\item
The total trace of the Einstein equations
\begin{eqnarray}
\label{trace}
-4\Theta^2 -6\dot{\Theta}=
3\kappa^2\left(3P-\rho\right) + 12\left(f -\mathcal{G} f_{\mathcal{G}}\right) \nonumber\\
+\frac{8}{3}\Theta^3 f_{\cal{G}\cal{G}}\dot{\mathcal{G}}
+\frac{ 18}{\Theta}\mathcal{G} \dot{f}_{ \mathcal{G} } +8\Theta^2\ddot{f_{\mathcal{G}}},
\end{eqnarray}
\item
The conservation equation for standard matter
\begin{equation}\label{cons:perfect}
\dot{\rho}=-\Theta\left(\rho+P\right).
\end{equation}
\end{itemize}
For FLRW spacetimes, the Gauss-Bonnet term is given by
\begin{equation}
\mathcal{G}=\frac{8}{9}\Theta^2\left[\dot{\Theta}+\frac{1}{3}\Theta^2\right]
=24\frac{\dot{a}^2\ddot{a}}{a^3}.
\label{gb}
\end{equation}
We can easily see that accelerating models have $\mathcal{G}>0$, while decelerating models have $\mathcal{G}<0$. In particular, any expansion history evolving from deceleration to acceleration must pass through $\mathcal{G}=0$. This observation will be of importance in the following discussion.
\subsection{Requirements for the existence of power--law solutions}
Let us now assume there exists an {\it exact} power--law solution to the field equations,
i.e., the scale factor behaves as
\begin{equation}
a(t)=a_0t^m\,.
\label{a-pl}
\end{equation}
From now on, we assume that $m$ is a \emph{fixed} real number.
If $0<m<1$, then the required power--law solution is \emph{decelerating}, while for $m>1$ it is
\emph{accelerating}. Since we know that within the standard paradigm, the expansion
history of the universe underwent a power--law decelerating phase, it is important to
study these kinds of exact solutions in our modified gravity models.

We further assume that matter can be described by a barotropic
perfect fluid such that $P=w\rho$ with $w\in[-1,1]$.
From the energy conservation equation, we obtain
\begin{equation}
\rho(t)=\rho_0t^{-3 m (1 + w)},
\end{equation}
and the Gauss-Bonnet term becomes
\begin{equation}
\mathcal{G}= 24m^3(m-1)t^{-4}  \equiv \alpha_m t^{-4}\;.
\label{G_pl}
\end{equation}
The negative sign of $\mathcal{G}$  for all decelerating models is reflected by $\alpha_m<0$ for the power--law models with $0<m<1$.

Using the background solutions above, we can write the Friedmann,
Raychaudhuri and trace equations in terms of functions of time $t$ only. We can assume with no loss of generality that $t>0$.
We then solve (\ref{G_pl}) for $t$ and re-write these equations in terms of
$\mathcal{G}$, $f(\mathcal{G})$ and its derivatives. The Friedmann equation for example becomes
\begin{equation}
f -  f_{\mathcal{G}} \mathcal{G} + 3m^2 \sqrt{\frac{\mathcal{G}}{\alpha_m}}
- \frac{96  f_{\mathcal{G}\mathcal{G}} \mathcal{G}^2 m^3}{\alpha_m}
- K\left(\frac{\mathcal{G}}{\alpha_m}\right)^{\sfrac{3}{4} m (1 + w)} =0\;,
\label{friedman-G}
\end{equation}
where $K=\rho_0/a_0^{3(1+w)}$. Note that for the power--law solution (\ref{a-pl}), $\mathcal{G}/\alpha_m$ is positive at all times by definition (\ref{G_pl}), and therefore equation (\ref{friedman-G}) is real-valued over the range of $\mathcal{G}$.

Since we want (\ref{a-pl}) to be a solution at all times, i.e. $\mathcal{G}$ spans over an entire branch of the real axis, we can interpret (\ref{friedman-G}) as a differential equation for the
function $f$ in  $\mathcal{G}$ space.
This differential equation has the general solution
\begin{equation}
f(\mathcal{G})=A_{m}  \sqrt{\tilde{\mathcal{G}}} + B_{mw}
\tilde{\mathcal{G}}^{\sfrac{3}{4} m (1 + w)} + C_1\mathcal{G}
+C_2\mathcal{G}^{\sfrac{1}{4} - \sfrac{m}{4}} \,.
\label{f(G)}
\end{equation}
We have abbreviated $\tilde{\mathcal{G}}\equiv 24\mathcal{G}/\alpha_m$, which we repeat is positive for the power--law solution (\ref{a-pl}), and
\begin{eqnarray}
A_{m}&=&-\sqrt{\frac{3}{2}} \frac{m^2(m-1)}{m+1} ,\\
B_{mw}&=&\frac{-2^{2 - \sfrac{9}{4} m (1 + w)} 3^{-\sfrac{3}{4} m (1 + w)} (m-1)
K}{4 - m (19 + 15 w) + 3 m^2 (4 + 7 w + 3 w^2)}
\end{eqnarray}
The constants $C_{1,2}$ are constants of integration. Since we know that the term linear in
$\mathcal{G}$ does not change the field equations, we can without any loss of generality
assume $C_1=0$. Furthermore, if we wish to ensure that for $m=2/[3(1+w)]$ and
$K=4/[3(1+w)^2]$, the theory reduces to GR, i.e., $f(\mathcal{G})=0$, then this constrains
$C_2=0$.
Hence, for an exact power law solution to exist, the required form of the function $f$ becomes
\begin{equation}
f(\mathcal{G})=A_{m} \sqrt{\tilde{\mathcal{G}}} + B_{mw}
\tilde{\mathcal{G}}^{\sfrac{3}{4} m (1 + w)} \,.
\label{f(G)1}
\end{equation}
We note that the above form of $f$ identically satisfies the other field equations,
if we similarly convert them as differential equations in $\mathcal{G}$ space.
The coefficients $A_m,B_{mw}$ are real-valued and non-zero unless $m=1$,
in which case $a(t)\propto t$.
In general, the function $f(\mathcal{G})$ is real-valued only if
${\mathcal{G}}/{\alpha_m}>0$,
which is true by construction for the exact power law solution  (\ref{a-pl}).
Similar results were found in \cite{Lidsey} using a scalar--tensor approach to $f(\mathcal{G})$--gravity.

It is interesting to note that an exact GR-like solution ($m=2/3(1+w)$) is possible with non-zero $C_2$ for $f(\mathcal{G})=
C_2\mathcal{G}^{\sfrac{1+3w}{4(1+w)}}$ for values of $w$ for which $f(\mathcal{G})$ is real--valued.

\subsection{Co-existence of decelerating power--law solutions and accelerating solutions}
To mimic the standard expansion history of the universe in $f(\mathcal{G})$--gravity, we assume there exists an exact decelerating power--law solution, and that the universe was well described by this solution in the past, before coming to the accelerated phase.  As argued in the previous section, the existence of the exact solution fixes the form of $f(\mathcal{G})$ as in (\ref{f(G)1}), and the deceleration fixes
the power $m$ as $0<m<1$, implying $\alpha_m<0$.

Now, if any \emph{additional} accelerating solution exists in the whole solution space of the model, then $\mathcal{G}>0$  for this
solution as evident from (\ref{gb}) . However,  we can see from the form of $f(\mathcal{G})$, that this is
not possible, as for $\mathcal{G}>0$ the function is no longer real valued.

This problem can be artificially remedied by including absolute values of $\tilde{\mathcal{G}}$
in (\ref{f(G)1}) (similarly to \cite{Zhou}), i.e. redefining
\begin{equation}
\tilde{f}(\mathcal{G})=A_{m} \sqrt{\left| \tilde{\mathcal{G}}\right|} +
B_{mw} \left| \tilde{\mathcal{G}}\right|^{\sfrac{3}{4} m (1 + w)}\,.
\label{f(G)2}
\end{equation}
This function $\tilde{f}(\mathcal{G})$ now seems to allow for both
a decelerating power--law solution and accelerating solutions. However,
$\tilde{f}(\mathcal{G})$ is not differentiable at $\mathcal{G}=0$ and hence no longer a
$C^2$-function, which is required for  any  Lagrangian in the Einstein Hilbert action. Any expansion
history evolving from deceleration to acceleration must pass through
$\mathcal{G}=0$, and we can easily see that the field equations are no longer defined at this
point. Hence we can conclude that  \emph{no well--defined $C^2$ action in
$f(\mathcal{G})$--gravity can allow for exact decelerating power--law
solutions to co-exist with accelerating solutions}.

\subsection{Comparison to $f(R)$--gravity}
It is interesting to note that in $f(R)$--gravity the same argument cannot be
carried through. One important difference between $f(R)$-- and $f(\mathcal{G})$--gravity
is that in $f(\mathcal{G})$--gravity the energy-momentum tensor decouples from the correction terms. In $f(R)$--gravity on the other hand, the correction term modifies the matter energy-momentum  tensor, which becomes rescaled to $\tilde{T}^M_{ab}=\frac{1}{f'(R)}T_{ab}^M$.  Consequently it is not possible to convert the field equations in $f(R)$--gravity to linear differential equations for the function $f(R)$ in
$R$ space, but instead one obtains a non-linear differential equation. This equation cannot be solved easily, and in particular does not have a unique solution as is the case in the $f(\mathcal{G})$ analogue. Additionally, since in $f(\mathcal{G})$--gravity matter decouples from the correction terms any matter dominated era must strictly obey the corresponding GR type power--law behavior.

The analogue also breaks down in the sense that any cosmologically viable model must make a transition from deceleration to acceleration,  and therefore go through $\mathcal{G}=0$ (see (\ref{gb})). This poses a problem for the simple models including terms of the form $\mathcal{G}^n$ with $n<1$, which are not differentiable at  $\mathcal{G}=0$.
The value of the Ricci scalar $R$ on the other hand does not necessarily pass through $R=0$, and therefore models including terms like $R^n$ may still be viable. In fact if $f(R)_{,R}|_{R\rightarrow 0\pm}=0$ for any $f(R)$--gravity model, then the plane $R=0$ is actually an invariant sub-manifold. This means that no solution can cross that plane in phase space. For example, for any $n<1$,  $R^n$ is not differentiable at $R=0$, but any solution can only approach this plane from either side and not reach it in finite time.
In fact, as shown in  \cite{cdct:dynsys05} and \cite{goheer07}, in $R^n$--gravity a Friedmann like matter dominated decelerated solution co-exists with a de--Sitter like accelerated solution. These two fixed points can be linked without $R$ changing sign, i.e. the field equations are well defined and fully differentiable along the entire orbit.

\section{Discussion and Conclusion}
The expansion history of the Universe is thought to have undergone a phase of decelerated power--law expansion followed by late time acceleration. Therefore, power--law solutions play an important role in cosmology as matter dominated phases that later connect to an accelerating phase.

In this work we have shown that \emph{exact} power--law solutions in  $f(\mathcal{G})$--gravity only exist
for the very special class of models given in (\ref{f(G)1}). In particular, many of the popular $f(\mathcal{G})$-models, e.g. all examples considered in \cite{felice08}, do not allow for any exact power--law solutions. This means that for these models, there exists no exact matter dominated solution, and therefore these models may not be of cosmological interest. 

Furthermore,
we have shown that decelerating power--law solutions cannot co--exist with \emph{any} accelerating
solutions, since no differentiable function $f(\mathcal{G})$ can admit both decelerating
power--law solutions and accelerating solutions. The problem may be
remedied by including absolute values in  (\ref{f(G)1}). However, this
makes $f(\mathcal{G})$ non-differentiable at $\mathcal{G}=0$, which is the value
$\mathcal{G}$ takes when the scale factor evolves from deceleration to acceleration,
and therefore cannot be ignored. This result seriously constrains the cosmological viability of $f(\mathcal{G})$--gravity,
since it may not be possible to obtain an expansion history similar to the  $\Lambda$\,-\,CDM model in this context.

This issue has not been addressed in recent papers \cite{Zhou}, where dynamical systems methods were used to study certain classes of $f(\mathcal{G})$--gravity models. It must be emphasized that even if equilibrium points corresponding to both decelerating power--law and accelerating solutions are found to co-exist as in \cite{Zhou}, there cannot exist any trajectories connecting these points, since these solutions can only co-exist for actions that are non-differentiable at the transition from deceleration to acceleration.

One possible way around this problem stems from the fact that our conclusions are based on the requirement that there exists an \emph{exact}
power--law solution. If however the scale factor behaves like $a(t)\propto e^{\Lambda t} t^m$, then at early times $a(t)\propto t^m$.
In this case the basic assumption of our analysis is not satisfied.

We therefore conclude that when looking for cosmologically viable solutions or fixed points in the dynamical systems state space in $f(\mathcal{G})$--gravity, one should not look for exact power--law solutions, but rather exact solutions that approximate decelerating power--law solutions at early times. Furthermore, we must restrict ourselves to functions $f(\mathcal{G})$ that are at least $C^2$ functions, and in particular differentiable at $\mathcal{G}=0$.

\acknowledgments
The authors would like to thank the National Research Foundation (South Africa) for financial support. RG would like to thank the University of Cape
Town for a Vice Chancellor's Postdoctoral Fellowship.

\end{document}